\documentclass{Interspeech}

\interspeechcameraready

\title{Geometric Analysis of Speech Representation Spaces: Topological Disentanglement and Confound Detection
}

\author[affiliation={1}]{Bipasha}{Kashyap}
\author[affiliation={1}]{Pubudu N.}{Pathirana}

\affiliation{Networked Sensing \& Biomedical Engineering (NSBE) Research lab, School of Engineering}{Deakin University}{Australia}
\email{b.kashyap@deakin.edu.au, pubudu@deakin.edu.au}

\usepackage{comment}
\usepackage{graphicx}
\usepackage{tikz}
\usepackage[activate]{microtype}
\begin{document}

\maketitle
\renewcommand\thefootnote{}
\footnotetext{Submitted to INTERSPEECH 2026.}
\renewcommand\thefootnote{\arabic{footnote}}

\begin{abstract}
Speech-based clinical tools are increasingly deployed in multilingual settings, yet whether pathological speech markers remain geometrically separable from accent variation remains unclear. Systems may misclassify healthy non-native speakers or miss pathology in multilingual patients. We propose a four-metric clustering framework to evaluate geometric disentanglement of emotional, linguistic, and pathological speech features across six corpora and eight dataset combinations. A consistent hierarchy emerges: emotional features form the tightest clusters (Silhouette 0.250), followed by pathological (0.141) and linguistic (0.077). Confound analysis shows pathological–linguistic overlap remains below 0.21, which is above the permutation null but bounded for clinical deployment. Trustworthiness analysis confirms embedding fidelity and robustness of the geometric conclusions. Our framework provides actionable guidelines for equitable and reliable speech health systems across diverse populations.
\end{abstract}


\section{Introduction}

The human voice simultaneously encodes emotional state, linguistic identity, and physiological health within a single acoustic signal. Separating these co-occurring information streams, a problem known as \textit{speech disentanglement}, is fundamental to clinical speech assessment~\cite{schuller2018speech}, voice conversion~\cite{qian2022content}, and speaker verification~\cite{li2023mutual}. While self-supervised models such as wav2vec~2.0~\cite{baevski2020wav2vec} and HuBERT~\cite{hsu2021hubert} yield impressive task performance, they produce opaque embeddings whose geometric structure remains poorly understood~\cite{pasad2023comparative}.

Understanding whether speech features occupy \textit{geometrically separable} regions of representation space remains an open problem. A diagnostic model requires that pathological and linguistic features reside in distinct manifold regions that support reliable decision boundaries, not merely that they capture different aspects of speech in aggregate. Without geometric characterisation, there is no guarantee that clinical features remain separable from accent-related variation across diverse populations.

The clinical urgency is concrete. Speech-based screening tools for Parkinson’s disease and dysarthria are entering healthcare systems that serve linguistically diverse populations~\cite{rusz2021guidelines,duffy2019motor}. Both Parkinson’s hypokinetic speech and non-native accents can produce reduced articulatory precision and imprecise consonant production~\cite{flege1995second}. If clinical features also respond to accent-related variation, healthy non-native speakers may be falsely flagged for neurological referral, while genuine pathology may be dismissed as accent variation in multilingual patients. Automated speech recognition systems already show significantly higher error rates for non-native speakers~\cite{koenecke2020racial}, and recent work on fairness in paralinguistic analysis~\cite{gorrostieta2022gender} underscores the broader pattern of demographic bias in speech AI.

We address this gap using four complementary clustering metrics applied to t-SNE embeddings of hand-crafted acoustic features. Our framework characterises \textit{where} speech dimensions reside in representation space and whether they form separable structures. Specifically, we:

\begin{enumerate}
    \item Propose a four-metric clustering framework (Silhouette, Davies–Bouldin, Calinski–Harabasz, Bootstrap Stability) for evaluating geometric disentanglement;
    \item Establish a manifold quality hierarchy (emotional $>$ pathological $>$ linguistic) consistent across eight corpus combinations;
    \item Demonstrate that hand-crafted clinical features maintain bounded geometric separation from accent features (overlap $< 0.21$), with confound severity quantified against a permutation null baseline;
    \item Validate embedding fidelity through trustworthiness analysis.
\end{enumerate}

\section{Related Work}

\subsection{Speech Representation Disentanglement}
Self-supervised speech models learn powerful but entangled representations~\cite{baevski2020wav2vec,hsu2021hubert}. Recent work has addressed pairwise separation: Cho et~al.~\cite{cho2025diemo} achieve cross-speaker emotion transfer through disentangled representations; Li et~al.~\cite{li2023mutual} apply MI-based decoupling for speaker verification; and Qian et~al.~\cite{qian2022content} propose content–style factorisation for voice conversion. Pasad et~al.~\cite{pasad2023comparative} provide a layer-wise analysis of what speech SSL models encode, revealing that different layers capture different information types. Wagner et~al.~\cite{wagner2023dawn} survey transformer-era emotion recognition and note that dimensional confounds remain poorly characterised. Despite this progress, systematic geometric analysis of three-way dimensional structure (emotional $\times$ linguistic $\times$ pathological) remains an open problem.

\begin{figure*}[t]
\centering
\begin{tikzpicture}
    \node[draw=gray, rounded corners=5pt, inner sep=1.5pt] {       \includegraphics[width=0.85\textwidth]{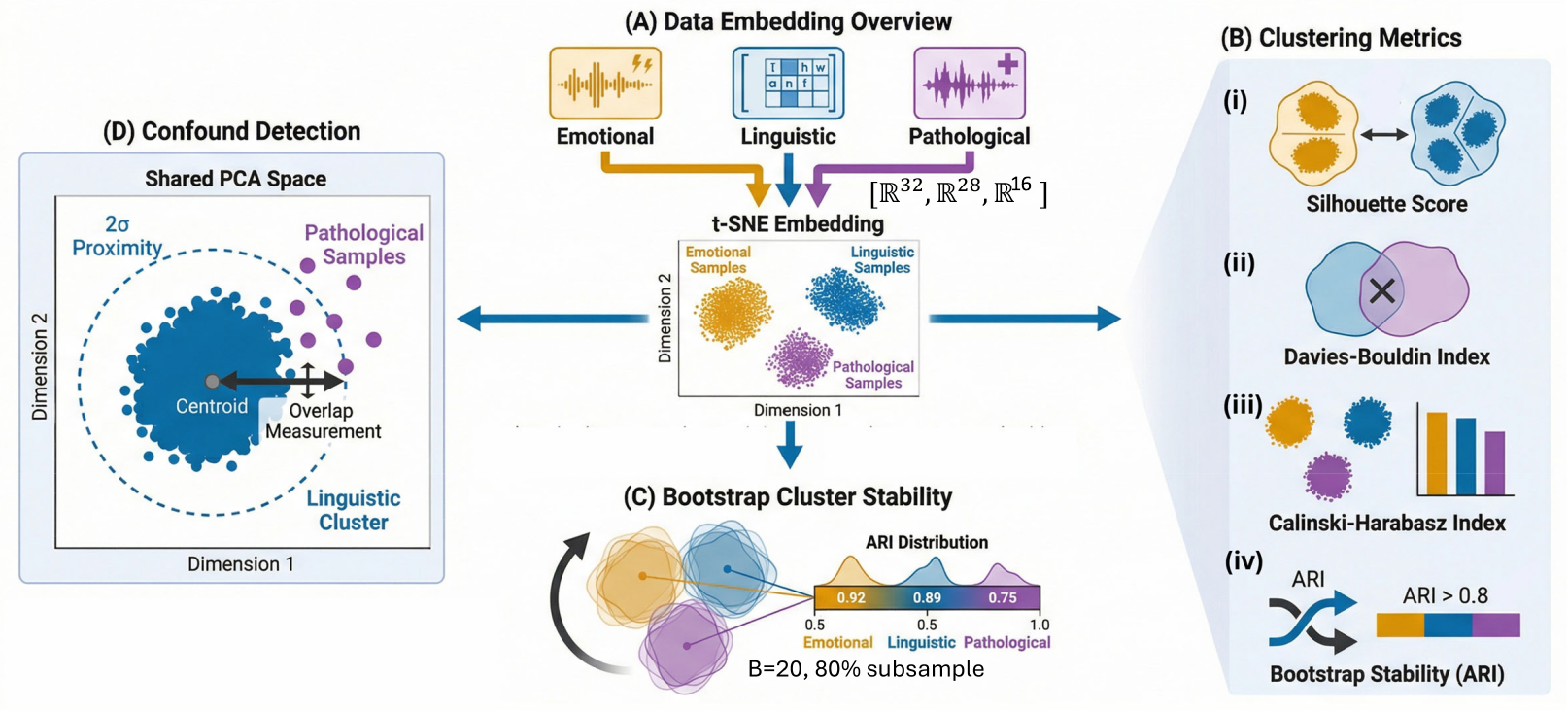}
    };
\end{tikzpicture}
\caption{Topological analysis framework. Overview: multi-dimensional features (emotional $\mathbb{R}^{28}$, linguistic $\mathbb{R}^{33}$, pathological $\mathbb{R}^{16}$) undergo t-SNE embedding, followed by three branches: (A) clustering quality via four metrics, (B) bootstrap stability ($B = 20$, 80\% subsampling), and (C) confound detection via $2\sigma$ overlap with a permutation null in shared PCA space.}
\label{fig:architecture}
\end{figure*}

\subsection{Clustering Quality Assessment}
Evaluating unsupervised cluster structure requires complementary metrics. The Silhouette Score~\cite{rousseeuw1987silhouettes} measures the sample-wise cohesion-to-separation ratio. The Davies–Bouldin Index~\cite{davies1979cluster} quantifies worst-case cluster overlap. The Calinski–Harabasz Index~\cite{calinski1974dendrite} captures between-to-within variance ratios. Bootstrap stability via the Adjusted Rand Index~\cite{hennig2007cluster} assesses robustness to subsampling. Von Luxburg~\cite{von2010clustering} establishes that no single metric suffices; we therefore combine all four.

\subsection{Clinical Speech Assessment in Multilingual Settings}
Automated assessment of Parkinson’s disease~\cite{rusz2021guidelines} and dysarthria~\cite{kim2008dysarthric} has shown clinical promise, but confounds with non-native accents have received limited attention~\cite{duffy2019motor}. Flege’s Speech Learning Model~\cite{flege1995second} predicts that non-native speakers produce articulatory patterns intermediate between L1 and L2 targets—patterns that may overlap with motor speech disorders. Koenecke et~al.~\cite{koenecke2020racial} document systematic bias in speech AI across demographic groups, motivating rigorous confound analysis.

\section{Methodology}

\subsection{Feature Extraction}
We extract three structured feature sets grounded in the source–filter model of speech production~\cite{fant1970acoustic}, using Praat~\cite{boersma2001praat} and librosa~\cite{mcfee2015librosa}.

Source features characterise glottal excitation dynamics: fundamental frequency (F0) statistics (mean, standard deviation, range, median, Q1, Q3), jitter, shimmer, and harmonic-to-noise ratio (HNR). These primarily reflect laryngeal vibration properties modulated by emotional state and vocal health~\cite{teixeira2013vocal,scherer2003vocal}. Filter features represent vocal tract resonance characteristics: formant frequencies and bandwidths (F1–F3, B1–B3) estimated via Burg LPC, together with 13 MFCCs and their first-order temporal derivatives~\cite{davis1980comparison}. These capture articulatory shaping driven by linguistic content and motor control. 
\textbf{Emotional features} $\mathbf{e} \in \mathbb{R}^{28}$ extend source descriptors with energy contours (RMS mean, standard deviation, maximum) and spectral statistics (centroid, flux, roll-off), consistent with the eGeMAPS framework~\cite{eyben2016geneva}. 
\textbf{Linguistic features} $\mathbf{l} \in \mathbb{R}^{33}$ augment filter descriptors with delta–delta MFCCs and rhythm-related parameters (tempo, duration), reflecting phonetic and prosodic structure~\cite{huang2001spoken}. 
\textbf{Pathological features} $\mathbf{p} \in \mathbb{R}^{16}$ target clinically relevant markers, including perturbation measures (jitter, shimmer, HNR), formant stability (F1–F3 coefficient of variation), and F2 transition velocity~\cite{rusz2021guidelines,kent2000research,kashyap2020cepstrum}.

All features are z-score normalised within each corpus combination to mitigate cross-dataset scale differences.

\subsection{Manifold Learning via t-SNE}
We employ t-SNE~\cite{vandermaaten2008tsne} to project high-dimensional feature vectors $\{\mathbf{x}_i\}_{i=1}^{N} \subset \mathbb{R}^{d}$ into two dimensions $\{\mathbf{y}_i\}_{i=1}^{N} \subset \mathbb{R}^{2}$ by minimising the Kullback–Leibler divergence between pairwise similarity distributions:

\begin{equation}
\min_{\mathbf{Y}} \mathrm{KL}(P \,\Vert\, Q)
= \sum_{i=1}^{N}
  \sum_{\substack{j=1 \\ j \ne i}}^{N}
  p_{ij} \log \frac{p_{ij}}{q_{ij}} .
\end{equation}

\begin{equation}
\sum_{i \ne j} p_{ij} = 1,
\qquad
\sum_{i \ne j} q_{ij} = 1,
\end{equation}

where $\mathbf{Y} = [\mathbf{y}_1, \dots, \mathbf{y}_N]$ denotes the low-dimensional embedding.  
$P$ and $Q$ are joint probability distributions over sample pairs in the high- and low-dimensional spaces, respectively: $p_{ij}$ is computed using a Gaussian kernel, while $q_{ij}$ is computed using a Student-$t$ kernel with one degree of freedom.

t-SNE is selected over PCA due to the nonlinear manifold structure of speech features, and over UMAP due to its stronger emphasis on local neighbourhood preservation, aligning with our objective of assessing cluster separability.

Parameters: perplexity = 30, learning rate = auto, 1000 iterations, PCA initialisation.

\subsection{Clustering Quality Metrics}
Cluster separability is quantified using three complementary indices.

\noindent\textbf{Silhouette Score}~\cite{rousseeuw1987silhouettes} evaluates per-sample assignment quality:

\begin{equation}
s(i) = \frac{b(i) - a(i)}{\max(a(i), b(i))},
\end{equation}

where $a(i)$ is the mean intra-cluster distance and $b(i)$ is the mean distance to the nearest neighbouring cluster.  
$s(i) \in [-1,1]$, with higher values indicating stronger separation.

\noindent\textbf{Davies–Bouldin Index}~\cite{davies1979cluster} measures average cluster similarity:

\begin{equation}
\text{DB} = \frac{1}{k} \sum_{i=1}^{k}
\max_{j \neq i}
\left(
\frac{\sigma_i + \sigma_j}
{d(\mathbf{c}_i, \mathbf{c}_j)}
\right),
\end{equation}

where $k$ is the number of clusters, $\sigma_i$ denotes within-cluster scatter, and $d(\mathbf{c}_i, \mathbf{c}_j)$ is the Euclidean distance between centroids. Lower values indicate improved separation.

\noindent\textbf{Calinski–Harabasz Index}~\cite{calinski1974dendrite} evaluates global cluster structure:

\begin{equation}
\text{CH} =
\frac{SS_B / (k-1)}
{SS_W / (n-k)},
\end{equation}

where $SS_B$ and $SS_W$ denote between- and within-cluster sums of squares, respectively; $k$ is the number of clusters, and $n$ is the total number of samples. Higher values indicate more compact and well-separated clusters.

\begin{figure*}[t]
\centering
\begin{tikzpicture}
    \node[draw=gray, rounded corners=5pt, inner sep=1.5pt] {
        \includegraphics[width=\textwidth]{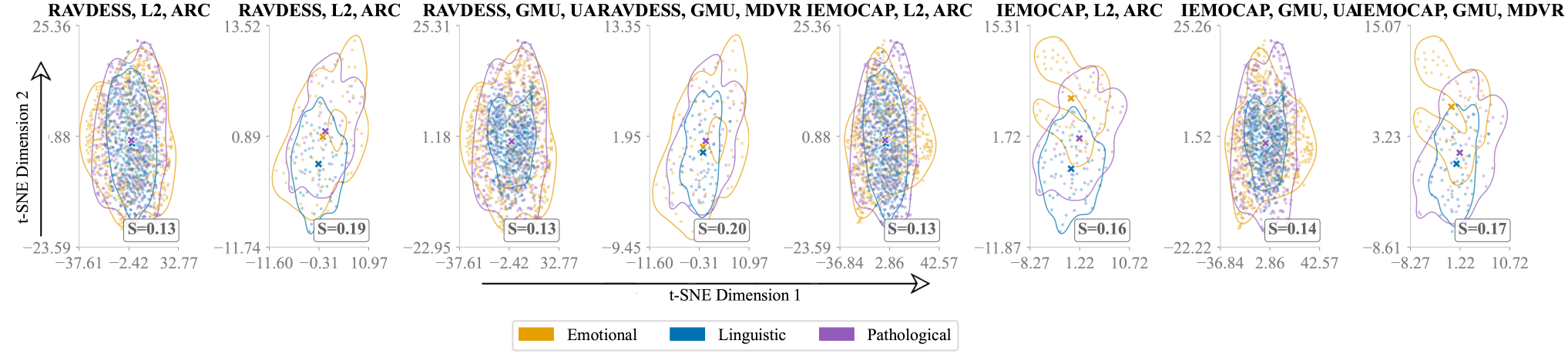}
    };
\end{tikzpicture}
\caption{t-SNE embeddings ($\mathbb{R}^2$, perplexity = 30) across all eight combinations. Per-dimension Gaussian kernel density contour~\cite{silverman1986density} (bandwidth=0.4, 30\% maximum density isoline) highlight manifold extent; filled regions show core density. Silhouette score badges (lower right) quantify per-panel clustering quality. The emotional-dominant hierarchy is consistent across all combinations, with pathological features forming intermediate clusters and linguistic features showing the most diffuse structure.}
\label{fig:tsne}
\end{figure*}

\noindent\textbf{Bootstrap Stability}~\cite{hennig2007cluster} assesses robustness to data perturbation:
\begin{equation}
\text{Stability} = \frac{1}{B} \sum_{b=1}^{B} \text{ARI}(\mathcal{C}_{\text{full}}, \mathcal{C}_b)
\end{equation}
where $B$ is the number of bootstrap iterations, $\mathcal{C}_{\text{full}}$ is the clustering obtained from all samples, $\mathcal{C}_b$ is the clustering from bootstrap sample $b$, and ARI is the Adjusted Rand Index measuring agreement between two partitions. We use $B = 20$ with 80\% subsampling.

For all metrics, we use KMeans with $k = 3$ clusters, matching the three semantic dimensions.

\subsection{Confound Detection}

To quantify pathological–linguistic geometric overlap, we compute:
\begin{equation}
\text{Overlap}(P_i, L_j) = \frac{|\{\mathbf{x} \in P_i : d(\mathbf{x}, \boldsymbol{\mu}_{L_j}) < 2\sigma_{L_j}\}|}{|P_i|}
\end{equation}
where $P_i$ is the set of pathological samples, $L_j$ is linguistic cluster $j$, $\boldsymbol{\mu}_{L_j}$ is the centroid of $L_j$, $\sigma_{L_j}$ is the mean standard deviation across dimensions within $L_j$, and $d(\cdot,\cdot)$ is the Euclidean distance. Because pathological ($\mathbb{R}^{16}$) and linguistic ($\mathbb{R}^{33}$) features differ in dimensionality, both are first projected into a shared PCA subspace of $d = \min(d_{\text{path}}, d_{\text{ling}}, 10)$ dimensions.

To interpret overlap magnitude, we compute a permutation null~\cite{good2005permutation} by pooling both feature sets, randomly reassigning labels ($n_{\text{perm}} = 200$), and recomputing Eq.~6. The 5th–95th percentile of this null distribution provides an empirical baseline for distinguishing genuine confound from chance proximity.

\section{Experimental Setup}

\subsection{Datasets}
We evaluate across six corpora. Abbreviations are used throughout tables and figures:

\textbf{Emotional:} RAVDESS~\cite{livingstone2018ravdess} (\textbf{RAV}; 1,440 utterances, 24 actors) and IEMOCAP~\cite{busso2008iemocap} (\textbf{IEM}; 10,039 utterances).
\textbf{Linguistic:} L2-ARCTIC~\cite{zhao2018l2arctic} (\textbf{L2A}; 24 non-native speakers, 6 L1 backgrounds) and GMU Speech Accent Archive (\textbf{GMU}; 2,140 speakers, 177 L1 backgrounds). 
\textbf{Pathological:} UA-Speech~\cite{kim2008dysarthric} (\textbf{UAS}; 15 dysarthric speakers) and MDVR-KCL (\textbf{MDV}; Parkinson’s, mobile recordings).

We analyse all 8 combinations ($2 \times 2 \times 2$) to ensure cross-corpus generalisability.

\subsection{Implementation}
Features are z-score normalised per combination before t-SNE projection. Clustering uses scikit-learn~\cite{pedregosa2011scikit} KMeans ($k = 3$, $n_{\text{init}} = 10$). Bootstrap: $B = 20$, 80\% subsampling. Confound overlap: $2\sigma$ in shared PCA space. Trustworthiness~\cite{venna2006local} is computed with $k = 15$ neighbours to verify embedding fidelity. 
\vspace{-2mm}

\section{Results and Discussion}

\subsection{Per-Dimension Clustering Quality}

Table~\ref{tab:clustering} presents clustering metrics aggregated across all eight combinations. A consistent hierarchy emerges: emotional features achieve the highest Silhouette Score ($0.250 \pm 0.057$), followed by pathological ($0.141 \pm 0.012$) and linguistic ($0.077 \pm 0.016$). All three metrics corroborate this ordering.

Figure~\ref{fig:tsne} visualises the manifold structure. Emotional features tend to form more compact clusters, whereas linguistic features exhibit comparatively more diffuse distributions. Per-panel Silhouette badges confirm cross-corpus consistency. Each badge represents the mean Silhouette score across all available dimensions within that subplot’s combination (Emotional, Linguistic, and Pathological).

The moderate absolute values ($< 0.30$) reflect the fact that feature sets share components by design (e.g., formants appear in both linguistic and pathological sets). The \textit{relative} ordering, rather than the absolute magnitude, constitutes the principal scientific finding. This consistent hierarchy (emotional $>$ pathological $>$ linguistic) has direct implications for system design: tight emotional clustering supports categorical classification; pathological features occupy an intermediate position, consistent with motor speech severity existing on a continuum~\cite{kent2000research}, suggesting regression-based architectures for clinical assessment; diffuse linguistic structure aligns with the combinatorial nature of phonetic variation across diverse language backgrounds.

\begin{table}[t]
\centering
\caption{Per-dimension clustering quality (mean $\pm$ SD across 8 combinations). Higher Silhouette and CH values indicate better clustering; lower DB values indicate better separation. Feature overlap across sets is by design (e.g., formants in both linguistic and pathological sets) and contributes to moderate absolute Silhouette values.}
\label{tab:clustering}
\vspace{-1mm}
\resizebox{\columnwidth}{!}{%
\begin{tabular}{lccc}
\toprule
\textbf{Dimension} & \textbf{Silhouette $\uparrow$} & \textbf{DB Index $\downarrow$} & \textbf{CH Index $\uparrow$} \\
\midrule
Emotional ($\mathbb{R}^{28}$) & $0.250 \pm 0.057$ & $1.448 \pm 0.161$ & $91 \pm 70$ \\
Pathological ($\mathbb{R}^{16}$) & $0.141 \pm 0.012$ & $1.859 \pm 0.212$ & $44 \pm 32$ \\
Linguistic ($\mathbb{R}^{33}$) & $0.077 \pm 0.016$ & $2.665 \pm 0.364$ & $22 \pm 17$ \\
\bottomrule
\end{tabular}%
}
\vspace{-3mm}
\end{table}

\subsection{Bootstrap Stability and Robustness}

Emotional clusters achieve the highest bootstrap stability (ARI: $0.82 \pm 0.08$), followed by pathological ($0.64 \pm 0.18$) and linguistic ($0.51 \pm 0.20$), mirroring the Silhouette hierarchy. The Pearson correlation between Silhouette and stability is $r = 0.74$ ($p < 0.001$), confirming that better-separated clusters are also more robust to subsampling.

Table~\ref{tab:trust} presents t-SNE trustworthiness. All values exceed 0.79, with emotional embeddings highest ($0.912 \pm 0.043$), followed by pathological ($0.876 \pm 0.050$) and linguistic ($0.809 \pm 0.007$). The uniformly high trustworthiness confirms that clustering metrics computed on t-SNE embeddings reflect genuine high-dimensional structure rather than projection artefacts. Together, high trustworthiness and the Silhouette–stability correlation provide independent confirmation that geometric conclusions are reliable.

\begin{table}[t]
\centering
\caption{t-SNE trustworthiness ($k = 15$) per dimension and combination. Values indicate the proportion of high-dimensional neighbours preserved in 2D. All values exceed 0.79. The mean follows the Silhouette hierarchy: Emotional ($0.912$) $>$ Pathological ($0.876$) $>$ Linguistic ($0.809$).}
\label{tab:trust}
\vspace{-1mm}
\resizebox{\columnwidth}{!}{
\begin{tabular}{lccc}
\toprule
\textbf{Combination} & \textbf{Emotional} & \textbf{Linguistic} & \textbf{Pathological} \\
\midrule
RAV--L2A--UAS & 0.964 & 0.814 & 0.926 \\
RAV--L2A--MDV & 0.888 & 0.799 & 0.826 \\
RAV--GMU--UAS & 0.964 & 0.811 & 0.926 \\
RAV--GMU--MDV & 0.888 & 0.812 & 0.826 \\
IEM--L2A--UAS & 0.949 & 0.814 & 0.926 \\
IEM--L2A--MDV & 0.848 & 0.799 & 0.826 \\
IEM--GMU--UAS & 0.949 & 0.811 & 0.926 \\
IEM--GMU--MDV & 0.848 & 0.812 & 0.826 \\
\midrule
\textbf{Mean $\pm$ SD} & $\mathbf{0.912 \pm 0.043}$ & $\mathbf{0.809 \pm 0.007}$ & $\mathbf{0.876 \pm 0.050}$ \\
\bottomrule
\end{tabular}}
\vspace{-3mm}
\end{table}

\subsection{Confound Analysis and Feature Interaction}

\begin{figure}[h]
\centering
\begin{tikzpicture}
    \node[draw=gray, rounded corners=5pt, inner sep=1.5pt] {
        \includegraphics[width=\linewidth]{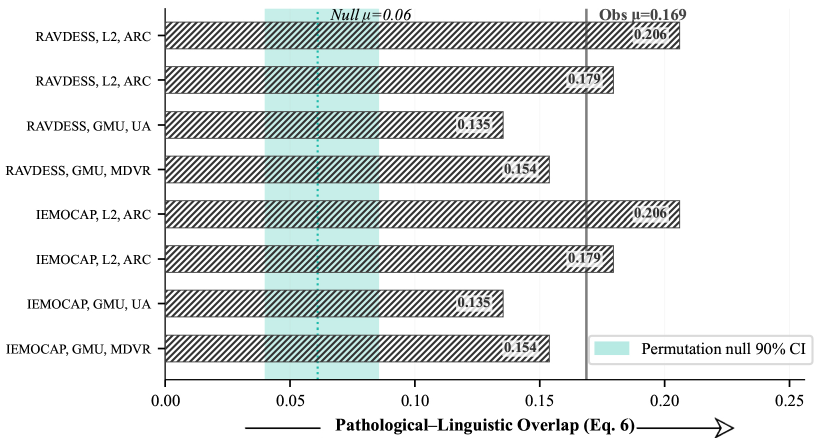}
    };
\end{tikzpicture}
\caption{Pathological–linguistic overlap (Eq.~6) across eight combinations, versus permutation null. The shaded region marks the 90\% confidence interval of the permutation null~\cite{good2005permutation} ($\mu_{\text{null}} \approx 0.06$, dotted). All observed values exceed the null baseline, confirming genuine shared structure, yet remain bounded ($< 0.21$). The solid line denotes the observed mean ($\mu_{\text{obs}} = 0.169$). L2A pairings show higher overlap than GMU pairings, suggesting that accent diversity improves separation.}
\label{fig:confound}
\end{figure}

Figure~\ref{fig:confound} presents pathological–linguistic overlap with permutation null comparison. Overlap ranges from $0.135$ (GMU pairings) to $0.206$ (L2A pairings). The permutation null baseline ($\mu_{\text{null}} \approx 0.06$) confirms that the observed overlap reflects genuine shared acoustic structure (e.g., formant components present in both feature sets) rather than chance proximity. Crucially, overlap remains bounded ($< 0.21$) despite intentional feature sharing.

The L2A–GMU difference (0.179–0.206 vs.\ 0.135–0.154) indicates that training data diversity itself functions as a confound mitigation strategy: the smaller L2A corpus concentrates accent variation near pathological feature space, whereas GMU distributes it more broadly. Overlap is identical across RAV and IEM pairings for matched linguistic–pathological combinations, confirming that the confound arises from linguistic–pathological feature interaction rather than emotional feature effects.

\subsection{Limitations}

The moderate absolute Silhouette values ($< 0.30$) indicate that none of the feature sets achieves sufficiently clean separation for reliable unsupervised classification. Although the relative hierarchy is robust, overall cluster separation remains partial rather than well-defined. The pathological datasets are comparatively small (UAS: 15 speakers; MDV: mobile recordings), and the narrow confidence intervals ($\pm 0.012$) may therefore reflect limited within-class variability rather than strong structural stability.

Feature definitions follow established conventions~\cite{schuller2018speech,eyben2016geneva}, but they were not explicitly optimised for geometric separability. Future work should assess whether learned representations~\cite{baevski2020wav2vec,hsu2021hubert} yield stronger separation. Finally, the PCA-based confound measure captures only linear structure and may overlook non-linear interactions; kernel-based alternatives could offer more sensitive detection of subtle dependencies.

\section{Conclusion}

Speech-based clinical tools require geometric separability of pathological features from accent variation. We presented a four-metric clustering framework evaluated across six corpora and eight combinations, establishing a consistent hierarchy: emotional features form the tightest clusters (Silhouette: $0.250 \pm 0.057$), followed by pathological ($0.141 \pm 0.012$) and linguistic ($0.077 \pm 0.016$). Confound analysis shows that pathological–linguistic overlap remains below 0.21, above the permutation null but bounded for clinical deployment. Trustworthiness analysis ($> 0.80$) confirms embedding fidelity. Our framework provides actionable guidelines for equitable speech health systems. Future work will extend this analysis to learned neural representations and broader clinical populations.


\bibliographystyle{IEEEtran}
\bibliography{mybib}

\end{document}